# A Policy Compass for Ecological Economics

Michèle Friend, Associate Professor
Department of Philosophy
George Washington University
Washington D.C.
Michele@gwu.edu

## Abstract

A policy compass indicates the direction in which an institution is going in terms of three general qualities. The three qualities are: suppression, harmony and passion.

Any formal institution[1] can develop a policy compass to examine the discrepancy between what the institution would like to do (suggested in its mandate) and the actual performance and situation it finds itself in. The latter is determined through an aggregation of statistical data and facts. These are made robust and stable using meta-requirements[2] of convergence.

Here, I present a version of the compass adapted to embed the central ideas of ecological economics: that society is dependent on the environment, and that economic activity is dependent on society; that we live in a world subject to at least the first two laws of thermodynamics; that the planet we live on is limited in space and resources; that some of our practices have harmful and irreversible consequences on the natural environment; that there are values other than value in exchange, such as intrinsic value and use value.

In this paper, I explain how to construct a policy compass in general. This is followed by the adaptation for ecological economics. The policy compass is original, and so is the adaptation. The compass is inspired by the work of Anthony Friend, Rob Hoffman, Satish Kumar, Georgescu-Roegen, Stanislav Schmelev, Peter Söderbaum and Arild Vatn. In the conclusion, I discuss the accompanying conception of sustainability.

Key words: Policy compass, policy decisions, qualitative accounting, ecological economics, environment, sustainability

---

[1] An institution is any of: a convention or habit, a norm or valued social practice and a formal institution. The latter has a formal structure and distinguishes itself from the other two by including explicit rules and a mechanism for re-enforcement or correction. The legal system, universities, banks, city councils, hospitals, libraries and so on are all formal institutions. But the compass can be used for less formal institutions, such as a norm or a practice.

# A Policy Compass for Ecological Economics

Michèle Friend, Associate Professor
Department of Philosophy
George Washington University
Washington D.C.
Michele@gwu.edu

## 0. Introduction

Policy decisions are largely made in terms of money – for maximising profit. This is appropriate when maximising profit in the short term is the highest or only consideration. At best, this is suitable for profit-only businesses, or for institutions in financial crisis, although even this is disputable (Varoufakis 2017). It is inappropriate on other occasions. Unfortunately, policy decisions are still, too often, made on a profit-making basis despite the fact that what goes into the calculation is very superficial.

While I hesitate to offer a diagnosis, I suspect that we make policy decisions in this way because we have one numerical figure: a money amount. This is thought to be simple, and we believe we understand it. Moreover, we think that if our institution has surplus money, then we cut ourselves the slack to execute the real mandate better. In other words, all too often, we *defer* trying to realise the mandate *directly* under two sorts of pressure, one is that it is too complicated to explain or understand the implications of non-monetary policy, the other is that at a later date, when we have the cash, we can think at greater leisure how to better realise the policy and what that means. So, even when the mandate is clearly not monetary, policy decisions are made in monetary terms.

This is no accident. In the modern world, our acceptance of finance-based decisions is systemic. In the modern world, we are taught from a very young age to behave as *homo-economicus*, and that institutions are better off if we make similar sorts of decision for them. Universities increasingly teach neoclassical economic theory and no alternatives (Söderbaum, 2017, 26) We also think that we understand credit and debit, which we do to some extent, but as a 'value' it is highly abstract and only reflects value in exchange. It follows that in many instances making policy decisions based the idea of maximising profit is inadequate.

As a policy maker, we could be more sophisticated. We *could* use some of the many lovely tools for making policy decisions: have recourse to multi-criteria decision aides as found in Shmelev (2012). Fitoussi, Sen and Stiglitz, propose a "dash-board" of such decision aides (Fitoussi et. al. 2010). See figure 1. But, unless one is trained to read these representations of data, or to read a table of data, it is very difficult to use them to make a policy decision. It is even harder to justify that decision to people who lack the training. One could defer to the authority of an expert trained in using such aides, but then we compromise democracy which is important in some institutions, and is recognised to be important when making decisions that affect the natural environment. When we defer to an expert, we have a technocracy, not a democracy. (Söderbaum 2017, 35).

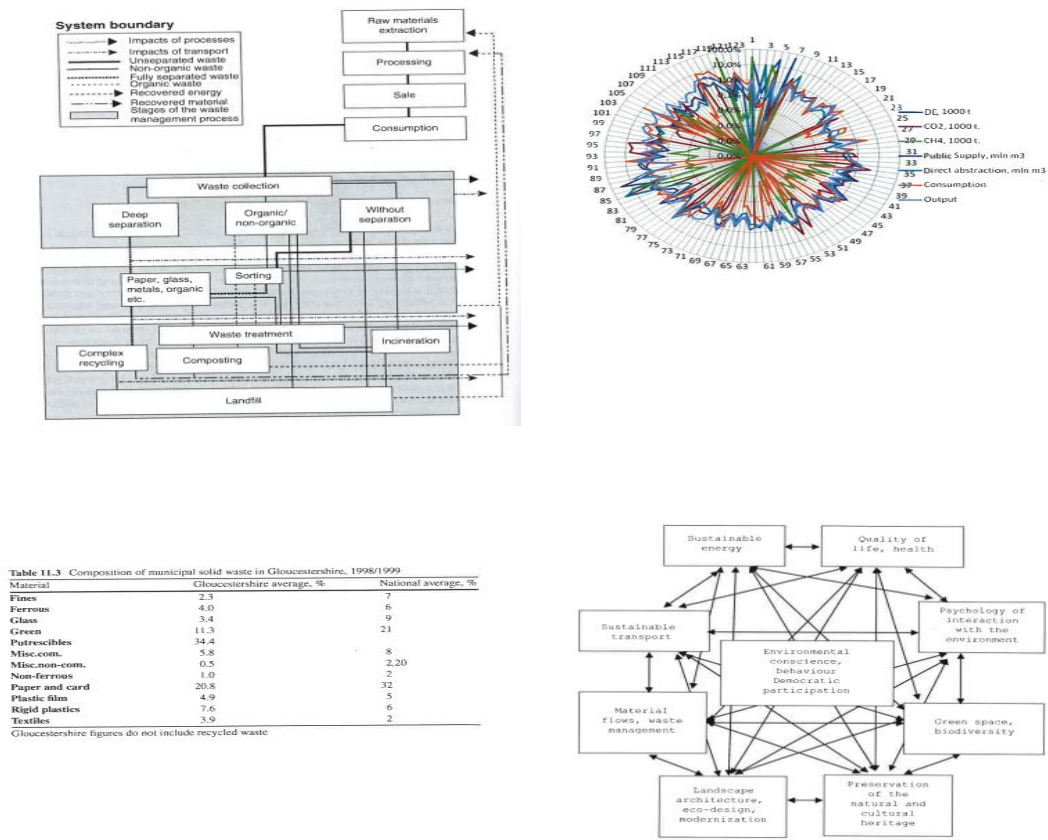

Figure 1: Multi-criteria decision aides.

In this paper, I propose a *new* tool for policy analysis, justification, development and change. I call it a 'policy compass'. It can be used for policies in any institution. Here, I adapt the policy compass to reflect the very important ideas in ecological economics: that the economy is a subset of society, and society is a subset of the natural environment. Here, 'subset' is an existential dependence relation. In developing the compass for an institution, in the adapted version presented here, we also incorporate the ideas that we live in a world subject to at least the first two laws of thermodynamics; that the planet we live on is limited in space and resources; that some of our practices have harmful and irreversible consequences on the natural environment (Rapport and Maffi 2010); that there are values other than value in exchange, such as intrinsic value and use value.

In contrast to the tools depicted in figure 1, visually, the policy compass is something very simple and intuitive. See figure 2. The simplicity of representation meets the demands of policy makers (Söderbaum, 2000, 54). After weighing and assessing individual statistics and facts, checking for robustness, we aggregate them to make one arrow on a trisected circle. The three sectors are: suppression, harmony and passion. The constructed arrow indicates the *direction* in which the institution is heading. The final arrow is in one of sectors, reflecting the fact that the statistics, when aggregated, show that *overall*, the institution displays this quality more than the others. The angle of the arrow *within* the quality indicates the degree to which it approaches, or tends away from, the other qualities. The length indicates the strength with which it sits in that quality. A shorter arrow would indicate more balance between the three qualities.

Behind the simple final representation lies a culturally sensitive, statistically robust and holistic construction.

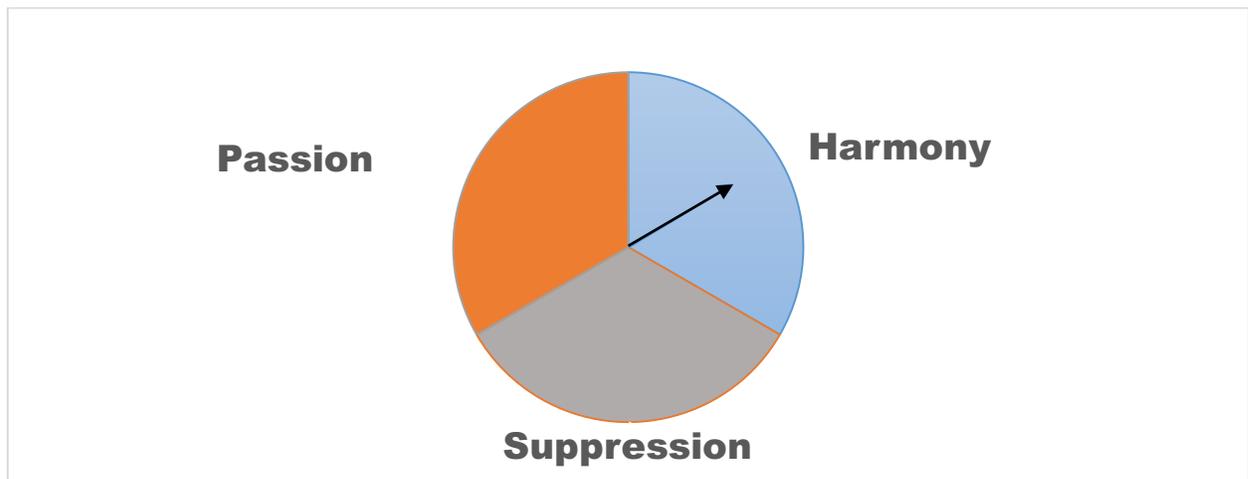

Figure 2: A Policy Compass.

Following Kumar (2007), the general qualities are inspired by the three *gunas* of Hindu, Jain and Buddhist philosophy: *tamas*, *raja*, and *sattva*.[3] I translate these as: suppression, passion and harmony, respectively. These are general, in the sense that other qualities fall under them. With the arrow, we can then make, analyse, critique and adjust decisions about the general qualitative direction of the institution, based on the angle and length of the arrow. As an institution, we might want to lengthen or shorten the arrow or we might want to change its sector or angle within a sector.

---

[3] The idea of contrasting these qualities is present in several deep religious traditions. I am *not* engaging the traditions as such. In particular, I distance myself from particular religious conceptions of when and how to apply them and how they interact with each other. I simply follow Kumar and make use of three qualities that can be applied to any institution. The meta-analysis of what to do when a particular quality is present and dominating will depend on the institution and it's cultural setting.

We construct the arrow by assembling statistical data in a table, making an analysis on the basis of the three qualities. The analysis is converted into a representation on the compass. Each statistic in the table is represented as an 'indicator arrow'. Lastly, we use a mathematical aggregation technique on indicator arrows to represent the *overall* aggregation of the information on the table. This represents the whole table at once. The statistics are chosen on the basis of: indicating the general quality, accuracy and availability and, in this adaptation, to reflect fundamental ideas in ecological economics.

Policy decisions are then made on the basis of the 'final arrow', as depicted in figure 2. How we create a new policy, how we adapt or change a policy, how we analyse or criticise a policy, how we justify a policy will then depend on *uncovering* how it is that we constructed the arrow in the first place, *in esse*, we return to the table of statistical data.

It will become obvious that we can less or more superficial in our policy decisions. We can create, adapt, analyse, criticise or justify a policy in a *superficial* manner by addressing or gerrymandering the *representation* of the statistics that swings, lengthen or shorten the final arrow. Or, we can create, adapt, analyse, criticise or justify a policy in a *less superficial* manner by addressing *particular statistics* that help or hinder the direction or length of the arrow. Or, *more deeply*, we can look at the *underlying causes* of the dominant statistics. The soundness and longevity of a policy will depend on the depth of analysis.

The structure of the paper is given by the following sections: (1) getting an overall sense of the trisected compass and how the statistics fit. (2) The method: choosing particular statistics, assembling them in a table and representing them on the compass. We represent each statistic with an 'indicator arrow'. (3) The mathematical technique used to aggregate the indictor arrows to end up with one final arrow. (4) The adaptation of the method to reflect the conceptions and normative concerns of ecological economics. (5) The policy compass and what it helps us to understand about our conception of sustainability. (6) Philosophical reflections and conclusion.

## 1. Explanation of the General Qualities

We start with three very general qualities.[4] They are inspired by[5] the three *gunas* of ancient Hindu, Jain and Buddhist literature. The *gunas* are: *sattva*, *raja* and *tamas*. I translate them as: harmonious, passionate and suppressive, respectively. Each of these general qualities, has sub-qualities. Examples of sub-qualities of harmonious are: pure, good, constructive, respectful, pleasant, soft, easy, light, natural and seamless. Examples of sub-qualities of passion are: active, plush, lively, confused, regal, exciting, sensational, perfumed, exotic, brassy, colourful, showy and spectacular. Examples of sub-qualities of suppression are: darkness, destructive, harmful, painful, chaotic, stinky, abrasive, constricting, despotic, putrid, diseased, depressing, morbid, violent, invasive and violent.

---

[4] We might think of these are meta-qualities.
[5] The use of the *guna* terms is only meant as a conceptual inspiration. It is not meant as a spiritual exercise in Hindu philosophy. The choice of the concepts of the three *gunas* will be discussed in the conclusion.

The importance of the three general qualities is that on considered reflection, almost every object, event, institution or society will have one that predominates. Time has the three qualities. "Living in the here and now, acting spontaneously and unselfconsciously, responding to a situation as it is and seeing the present moment is sattvic." (Kumar, 27). "Dwelling on the future is rajasic [especially when] … we become involved in exciting projects." (Kumar 28). "Living in the past is tamasic: Why did you do that? Why didn't I do that? You shouldn't have done that! We complain and moan." (Kumar 29). Food has the three qualities: sweet and light food is sattvic, spicy food is rajasic and heavy and stale food is tamasic. (Kumar, 30). Buildings have these qualities. Houses are sattvic, palaces are rajasic, prisons are tamasic. (Kumar, 34).

Try the following exercise: compare a plastic flower, a flower in a field and a cultivated cut flower for decorating indoors. The plastic flower falls under: 'suppressive' because it is dead and cynical, is made of plastic which is made from fossil fuels, plastic flowers are often dusty. The flower in the field falls under: 'harmonious' since it is alive, natural, pleasant, plays a role in the local ecology, brightens the landscape, helps living organisms and so on. The cultivated cut flower falls under: 'passionate', since it is usually larger and more extravagant than natural flowers, it is cut and taken away from its natural surroundings, so no longer contributes to the ecology, but adds beauty to the household. The cultivation shows mastery over nature, an enhancement of nature.

Try another comparison; this time one that is more institutional. Compare a primary school, a prison and a world-class sports team. The primary school falls under: 'harmony'. This is because it is normal for children to receive a primary school education. This helps to unite the society, giving the students social skills and literacy in numbers and letters so that they can communicate more widely. The prison falls under: 'suppressive' since prisoners are restricted in their movements, have little control over their daily routine, are punished for disobeying rules and are constantly watched. A more enlightened prison aiming at re-integration in society will still be suppressive, but, provided that the techniques used are kind and constructive, its arrow will tend more towards harmony than that of a retributive prison. The world-class sports team falls under: 'passionate'. They show outstanding physical ability, are matched only by the best in the world, bring pride and excitement to those who follow their results and they might enjoy a high degree of publicity.

Outward appearances and first associations might be deceptive. A person might be very wealthy and live in a rich dwelling, so we would suppose that person to be *rajasic* or passionate, but spiritually, he, or she, might be serene and modest and so *sattvic* or harmonious. In contrast, a person might appear modest and unassuming, so appear *sattvic* or harmonious, but on a spiritual level be full of rage and greed, and so is more *tamasic*, or suppressive. The analysis for the entering data on the table is better if we are sensitive beyond mere appearance.

Not only might outer appearances be deceptive, but institutions and people change in their general qualities. For example, a school might change over time. It might start as harmonious in its first years, then it might move into the general quality of passionate as it gains a strong reputation and gains a reputation for academic success. In an attempt to protect the high reputation, the school might become suppressive: with more rules, high standards (so failing more students) and an increase in ruthless measures taken to remove students who disrupt the flow of teaching, and mar the reputation. So, general qualities of an institution can change over time. We can track such changes by looking at the statistics that separately indicate the general qualities at different times.

We use a compass to indicate the dominating quality. To represent this, we divide a circle into three equal parts with radii, and label each third with one of the general qualities. We draw arrows (indictor arrows or final arrow) to give us a qualitative 'compass reading'.[6] See figure 2.

We can take different attitudes towards what the arrow indicates, and we can be more or less subtle in our policy changes. In general, and in Ancient Indian philosophical thinking, *sattva*, or harmony is the path of wisdom. So we re-balance a person or an institution by guiding them more towards harmony. In Western and Modern thinking, where we seek indications of 'progress', we tend to be attracted to passion in an institution, but, there is the danger that passion falls towards suppression. For example, as school policy makers for our passionate school, we are under the threat of the arrow moving too much towards suppression. We then can make a choice about what attitude to take: to align policy with Ancient Indian Philosophy, or with more Modern thinking. This is a philosophical choice that we make, and we should do so consciously, that is conscious of our having made a choice.

More specifically, we might take a philosophical / holistic view and promote harmony, sacrificing some of the passion in order to re-balance the school away from suppression. Or, we might have a more *mercantile / commercial* view and want to preserve the high reputation at all (other) costs. How might we do this? We decide that the suppression is the price we have to pay for the high reputation. We then maintain or increase the suppressive measures. Or we might be superficial and think that the *appearance* of indicators of suppression should be addressed, but we want to do so in such a way as to not sacrifice the high reputation. For this, we have to be subtler. We can export (to other schools) the disruptive students, so no suppressive indicators show up in *our* own school – there are fewer detentions, punishments, failed students and so on. Or, more superficially, we can re-represent the statistical data that pulls the arrow into the 'suppressive' so that the final arrow has less length, or swings back into the passionate third. How superficial we want to be in our analysis depends on how deep we want our justification to be for policy changes or decisions, and on the longevity we wish for those decisions.

Claim 1: The deeper the analysis, *ceteris paribus,* the greater the longevity of the policy.

So, we can make, analyse, justify, criticise, modify policy based on the final arrow and our ambitions for the institution. This is the purpose of the policy compass. How do we construct it?

## 2. The Indicator Arrows: Choosing Statistical Data

The final arrow is the result of aggregating statistical data represented by indicator arrows. We shall see the mathematical method of aggregation in the next section. For now, let us concentrate on the statistical data. Each statistic is represented by an indicator arrow. We have several initial steps.

---

[6] Lest there be some confusion, let us be clear. The compass does not have North, South, East and West. Instead, it has three thirds. The general qualities give us a general qualitative orientation of our arrow.

As per any policy decision, we want to choose data that is available, accurate (recent, representative and sufficient) and which we can safely assume will be available in the future (if we are interested in comparing policy of one institution over time) or is available for comparative institutions (if we want to compare institutions to each other). Thus,

**Step 1**: our first task is to find some data on the institution.

**Step 2**: verify that the quality of the data is reasonably high.

Such tasks are not meant to be very controversial, at least in the 'Western'/'Northern' world, but might be much more challenging to carry out in some communities or for some institutions. The ease of carrying out the tasks depends on the formality and context of the institution. That is, more casual, impromptu 'institutions' will have less data: a spontaneous village meeting, a protest march, a holiday celebration... A more stable and formal institution, such as a government, a well-established industry or a university will all have relatively high quality data available for analysis.

**Step 3**: for any institution, and for the data available, categorise the data in terms of the general qualities.

This third step is not all that mysterious, but there are some complications to be seen soon. Starting with the banality of the step: when someone informs us of a statistic, he, or she, does so to indicate a general quality to which we are supposed to react emotionally. The person might elicit the general sensation: that by-and-large everything is running smoothly (harmony), or that we should be on our guard, feel angry, frustrated or want to take political action (suppression) or that we should feel excited, proud/ jealous or passionate. A borderline feeling such as 'alarm' might be thought of as part of suppression and tend more-or-less towards passion. To construct the compass, we want consider statistics that each *positively* indicate *one* of the three general qualities. By 'positively' I mean that the statistic indicates the *presence* of the general quality, not its absence. This is a bit confusing with suppression, since it is often thought to be a 'negative quality'.

To allay the confusion, distinguish between the numerical conception of 'positive' and the emotional or normative conception of 'positive'. For the indicator arrows we choose statistics that *numerically* positively represent suppression. If they very strongly represent suppression then they will have *greater* length. For example, negative (in our feeling) national statistics that will positively (in the numerical sense) show suppression is the number of: suicides, fatal accidents, crime rates, natural disasters or percentage of prisoners. If these are small, negligible or not alarming, then we still place the indicator in suppression, but we shorten the length of the arrow that represents the statistic on our compass.

This reflection on the qualitative and emotional reaction to the data draws a *normative* and *culturally sensitive* aspect to the analysis. This is deliberate, and is considered to be a *strength* of the analysis: that we can now be quite explicit as to where and in what sense our policy has a normative element, and that different cultures might react in different ways to the 'same' statistic, and might change their reaction over time. For example, what were felt to be alarming rates of death by motor car accident are not felt to be as alarming today!

If there is too much disagreement about which general quality a statistic belongs to, then drop that statistic and look for another, or split it into two statistice with details that make it clea which general quality it exemplifies. For example, we might have ot be explicit about the perspective from which a statistic is viewed: from this perspective the indicator indicates this, but from that perspective it indicates something else. We now have two indicators. The sub-culture sensitivity is then made explicit.

Let us look at more examples. National statistics are fairly accurate and available (now, in the future and in several nations). Statistics that indicate harmony in a nation might include: literacy rate, participation in an electoral process, longevity of the population (provided this is 'normal'). Statistics that numerically positively indicate the emotionally negative quality of suppression might include: percentage of people in prison, a large gap between 10% highest income earners and 10% lowest income earners, volumes of waste or pollution. Statistics that indicate passion might include: high level of education, sporting achievements, Nobel prizes, natural wonders, number and volume of festivals. Which statistics fall under which general quality varies with cultures and sub-cultures. We shall return to this in a moment.

It is not easy or obvious how to come up with these, and for emotional or psychological reasons, many statistics that are cited in the media, and so are readily available, are more naturally placed in the passionate and suppressive sectors. That is because we pay attention to them because suppression elicits high emotions, and as I wrote earlier, one of the reasons that someone cites a statistic is to draw our emotion-laden attention to something, and increasingly, the different media institutions are run as a business for profit, and cannot 'sell' the media without eliciting strong emotional reactions. More systemically, as 'consumers' of media, we have been conditioned to *expect* to have an emotional reaction when reading the newspaper, watching the news on the television and so on. And that has become our (not very good) reason for purchasing that form of media.

We can counter-balance the emotion by citing a (less publicised because emotionally boring) statistic that indicates harmony. Paying attention to the emotional quality of a statistic is part of the exercise. We are aware that the emotional quality varies with culture, individual psychology and individual past history, including being recently psychologically 'primed'. Finding statistics that indicate harmony might be more difficult, and it is unlikely that we can find them in the media.

Let us examine the cultural element more closely. How we react emotionally to some statistics, depends on our culture. Thus, one culture might view a statistic as indicative of the general quality of harmony while another might view it as suppression. Take for example, a low divorce rate. This might be due to harmony in the marriages in the society, since there is little need for divorce. In another society, we might think that the social norms and laws make it so difficult to divorce, or the economic structure makes it too dissuasive to divorce and so a low rate of divorce indicates the general quality of suppression in too many marriages. Therefore, the proffered name for the statistic is not always enough to tell us which third the statistic belongs to. It's significance (that the number is high) and its context are also important. Which general quality the low divorce rate indicates has to do with a wider context and general cultural values.

If we have difficulty here, we can do two things. One option is to drop the statistic. The latter is less helpful for the compass construction for reasons of robustness of the final arrow.[7] The alternative is to be more specific about the name for the statistic, splitting it into two, to reflect the different contexts or ways of thinking about the statistic. We qualify the name, or give it a longer more nuanced name. So, we might call it, for example, "divorce rate where there is no stigma attached to divorce and it is economically not dissuasive for either party", and "divorce rate where this is highly stigmatised" as opposed to just "rate of divorce". Re-naming adds to what I have called 'depth' of analysis, and will play into the justification for, and longevity of, policy decisions made on the basis of the compass.

Taking seriously these complications is what adds depth to our policies. Remember, we use the final arrow to make new policies, adapt or change policies, analyse or criticise policies or justify policies. And remember that we can do all of these things in a more or less *superficial* manner. But remember claim 1!

Summarising: the third task is delicate but also adds normativity and depth. Continuing with the method, we want to represent each statistic as an indicator arrow on the circle. The first task is sufficient for us to draw arrows at all. The second task ensures reliability. The third task situates each indicator arrow in one of the sectors. We draw a table with one column for general quality, a second column for the name and any qualifiers. What of length and direction? See table 1.

| General Quality | Name of Indicator and Notes | angle | length | Corrected length |
|---|---|---|---|---|
| **Harmony** | Stability in wages | 100 | .5 | .15 |
| **Harmony** | General good health of employees | 30 | .8 | .3 |
| **Harmony** | Natural zone in space | 110 | .2 | .012 |
| **Passion** | Money earning charity drive | 140 | .2 | .012 |
| **Passion** | Publicity for art show | 240 | .2 | .012 |
| **Passion** | Tree planting | 150 | .1 | .01 |
| **Suppression** | Money ill spent on equipment | 320 | .6 | .78 |
| **Suppression** | People fired from company | 270 | .6 | .78 |
| **Suppression** | Use of energy in building | 280 | .3 | .09 |

Table 1. An example of a table for plotting indicator arrows.

We now have to be even more sensitive to culture and pay close attention to the nuance surrounding the qualitative measure of the indicator arrow. Some statistics will fall in the middle of the third, and some will tend towards one of the other thirds, in limit cases an indicator arrow might sit right on the border between two qualities. In such a case, I recommend that we choose another statistic, for reasons concerning the aggregation technique in the next section. The angle, or direction of the indicator arrow within the third, will be represented by degree. If a statistic by-an-large indicates harmony, it is located in the harmony third. But the arrow does not have to be placed in the middle of the third. It might tend towards passion or towards suppression.

---

[7] Robustness will be addressed when we have constructed our sector arrow in the next section.

Let us take an example. Say, there is a high literacy rate in a country. This might be because the standard of living is sufficiently high throughout the population that every family can afford to send its children to school, and the schooling system is well run, in fact a high proportion of the population then goes on to higher education. There is, moreover, good support for children with disabilities that affect literacy, so even they gain a good proficiency in reading despite their learning difficulties, there is a good transportation infrastructure for getting children to school, they are well nourished and healthy, so that they have the energy and attention to learn and the leisure time after school to rest, play or re-enforce the lessons learned in the classroom and so on. This is quite impressive, and so the arrow will be in the third labelled "harmony" but will tend towards "passion". There is no precise degree that will represent this. But there are different protocols we can use to determine a particular angle. Suggestions will be made soon.

In contrast, the high literacy rate might have some sinister underlying causes. Illiterate children might be forcibly taken from their parents and put into unpleasant boarding schools, the alternative to not becoming literate might be very unpleasant – leading to poverty, poor work conditions, low self-esteem and so on. Or, in a very different situation, the literacy rate might be measured in terms of 'functional literacy' that is, just enough to read commands or directives, but most of the literate population does not enjoy the greater fruits of education, and do not go on to higher education, do not read novels or do not read well enough for written political engagement. In this case, the arrow still sits in the harmonious sector but tends towards suppression.

Now for a discussion of protocol to determine a more precise angle.

**Step 4**: For the indicator arrows, choose or make up a protocol to determine the precise *degree*, or angle.

One easy protocol is to use visual feedback. We display an indicator arrow on the circle and see how people feel it represents what they think, we move it around, change its length, discuss the changes until we reach consensus. It is surprising how quickly this can be done, and how easy it actually is.

An alternative protocol is to vote. Say, a large group decides on an indicator arrow for the suppressive third. We then vote on whether the arrow should then swing towards harmony or passion. If 20% would swing the arrow towards harmony, and 80% would swing it towards passion, then since there are 120 degrees allotted to suppression sector, the arrow will be $96^0$ towards passion.

We might want to do something more sophisticated than straight voting, and have degrees of vote, so people who feel strongly, medium or lightly about their choice. Or we might accord different weights to votes by different people: people more or less affected by the institution, or people with greater knowledge, for example. These are all parameters *worth* discussing and considering carefully, since they indirectly answer to claim 1. Eliciting these discussions and being explicit about the decisions made is a deliberate part of the exercise of constructing the compass. These questions are philosophical and important. But they also answer to robustness considerations which we shall see shortly.

When making policies, based on statistical findings, it is highly relevant and important to have these discussions about what it is that a statistic indicates in general, and not be afraid to change the name to reflect further subtleties. Such discussion might be long and frustrating for some people, but they will save time in the long run, due to the stability of the policy. The stability is promoted by understanding how others helping to design policy, or react to policy, think about the situation indicated by the statistic, and by the transparency of the procedure. Robustness will be further promoted by the sheer number, and variety of types, of people consulted.

What of length? This is just as important as degree, but in some ways it is less problematic. Length of arrow is a numerical measure that is used for comparison of that statistic with others of the same sort. Length will be nothing more mysterious than a function of scale, what we think are reasonable parameters and where other comparable statistics lie. To dispel the mystery, think of our usual representation of comparative statistics. We start by stating a fact. For example: the average longevity of the population is seventy years. This bald statement will elicit no emotional reaction without some context which might be known already or which might need to be articulated. Say that this is the statistic today, and that it was higher twenty years ago, when it was seventy-six. To represent the relationship between the two statistics we draw a graph with a bottom line labelled with a progression from left to right to indicate time, and another vertical line at the left labelled bottom to top to indicate measures of longevity. We usually will miss out the possibility that longevity is equal to 0 – 50 years. The maximum would be, say, 100 years. The dates will not start with the beginning of mankind, but might track longevity over the last 50 years. We then plot the two measures on the graph. We decide on the scale and parameters of representation almost without thinking, and deciding on the length of the arrow for the compass is similar.

However, there is a subtlety we should address immediately. We should not confuse the (a) exercise of fitting the representation aesthetically on a graph, and (b) gerrymandering the aesthetic representation it in such a way as to elicit or increase/ decrease an emotional reaction. Whether the representation elicits alarm or not will depend on choice of scale. See figure 3. On the left side, we have a small scale of zero to six, and the differences between the lines looks large. Had we chosen a scale of minus twenty to fifty, the three lines would flatten out and look much the same. Now consider first the aesthetics: that (a) we have to make the representation on a page of a book or article or screen (so this determines something of the parameters of possible scales) and we might think of future or past measures of longevity, not bothering with longevity below 40, since this is 'unimaginable'. These are features of representation with which we work every day. And this is just how we determine scale.

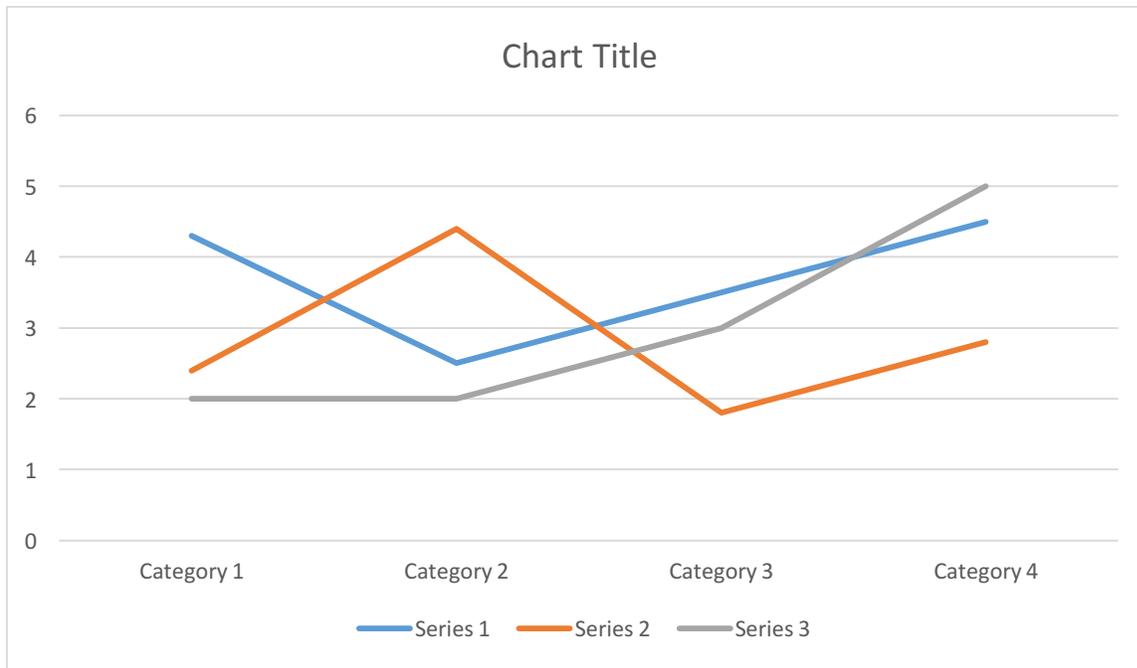

Figure 3: graph with chosen parameters along horizontal axis.

In the case of our indicator arrow on the circle, its length is then determined in the same way but adapted to a circle. *Ab initio*, the maximum length is the radius of the circle for an individual indicator arrow. For just one indicator arrow, we would fix the parameters according to what is reasonably imaginable. The average being half the radius. The statistic is then very high if the number reaches the edge of the circle but is low if it is close to the centre.

(b) We might want to work backwards and 'down-play' the emotion elicited. We then change the scale. This is not recommended for reasons of robustness, or objectivity which in turn depends on honesty in representation.

Now note: say longevity of a population belongs in harmony. A high longevity is more harmonious, say, than a low longevity, subject to considerations about what it is that is influencing the statistic be it for suppressive reasons (no laws allowing people to terminate their lives, so the quality of life is very low at the end of life) or for passionate reasons (the quality of health is very good even at the end of life). To re-enforce the quality of harmony, the arrow will be longer. A longer average longevity brings the arrow closer to the edge of the circle.

**Step 5**: For every indicator arrow, determine the length of the arrow following some protocol.

None of the above tasks is impossible, although for some analysts or policy makers it might be worth making the protocol explicit, again adding depth to the analysis. Developing the indicator arrows and accompanying protocols is the hardest, most time consuming and most qualitative and normative part of the exercise. However, for making future compasses, once the norm is in place, the discussions about protocol might not have to be revisited each time for each statistic. So, the time invested in deciding on protocol in the first place is well worth the effort in the future.

**Step 6**: Enter the statistics on a table with four columns: the general quality, the name and any qualifiers, the angle and the length.

Having chosen and developed the indicator arrows, we enter these in a table. Each arrow has a general quality, a name, an angle and a length. See table 1.

**Step 7**: Correct the length for the purposes of aggregation.

Note that in table 1 there are three indicator arrows for each sector. This is a very small number, and we would want more. Since there should be several arrows in each sector, to add robustness, and to prevent our sector (the aggregated arrow within a quality) growing out of proportion as we add more and more statistics. We also want to allow for a *different number* of statistics in each quality-sector. To accommodate this we have to do a little calculation. We want to emphasise the shorter arrows and flatten out the longer ones. We take the logarithm base two of the length and divide it by the number of statistics in that sector.[8]

The reason for making the correction is this. We want a nice representation for our final arrow. To stay within the bounds of the circle, we divide by the number of indicators in that sector. But not every indicator is as significant as the others. The shorter ones give us very little information, and might be added for reasons of robustness where we are looking for convergence of the sector arrow (to be explained later). These should not carry much weight. For the longer arrows, we want to gradually flatten them out so that they get close to the radius of the circle but not beyond. Long arrows are 'outliers'. The flattening out is calculated with the logarithm base two of the length. The calculations ensure that each indicator arrow's contribution to the aggregation should be in proportion to its importance, and the sector arrows do not shrink as we add more indicators to our table.

That is, there are significant indicators (with long length), sometimes called 'outliers'. There are insignificant ones (with short length). They tell us very little. To increase the robustness of sector arrows (ones within a general quality) we want as many arrows as possible. However, at some point, as we increase the number of indicator arrows, we start looking to less and less significant indicators, so the *aggregated sector arrow* would shrink as we added these insignificant arrows (because each is divided by the number of arrows in that sector). Under the robustness condition of adding more arrows, our final compass arrows would all converge to the dot in the middle of the circle. Instead we want the sector arrows to converge to a stable length for that sector as we add more indicator arrows. This will become clear once we see more of the compass construction, and the technique of aggregation.

---

[8] This will not make much sense until the end of section three, so the reader might want to return to this after reading sections three.

## 3. The Mathematics of Aggregating the Indicator Arrows

The mathematical modelling has ten steps.

**(i)** Plot each indicator arrow in its particular third with the corrected length. Plot them head to tail. The first has its tail at the centre of the circle. The second has its tail at the head of the first and so on. See figure 4.

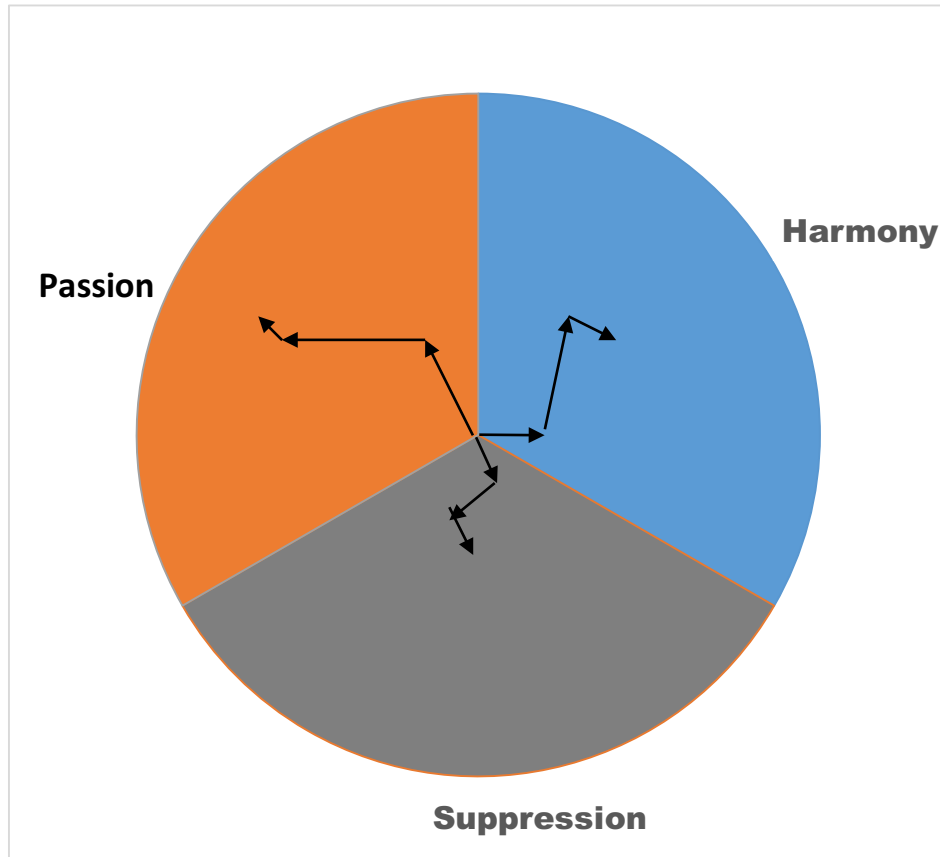

Figure 4: indicator arrows plotted tail to head.

**(ii)** Within each third, draw a 'sector arrow' from the centre of the circle to the head of the last arrow in that sector. Sector give us an overall-reading of that general quality for the institution. See figure 5.

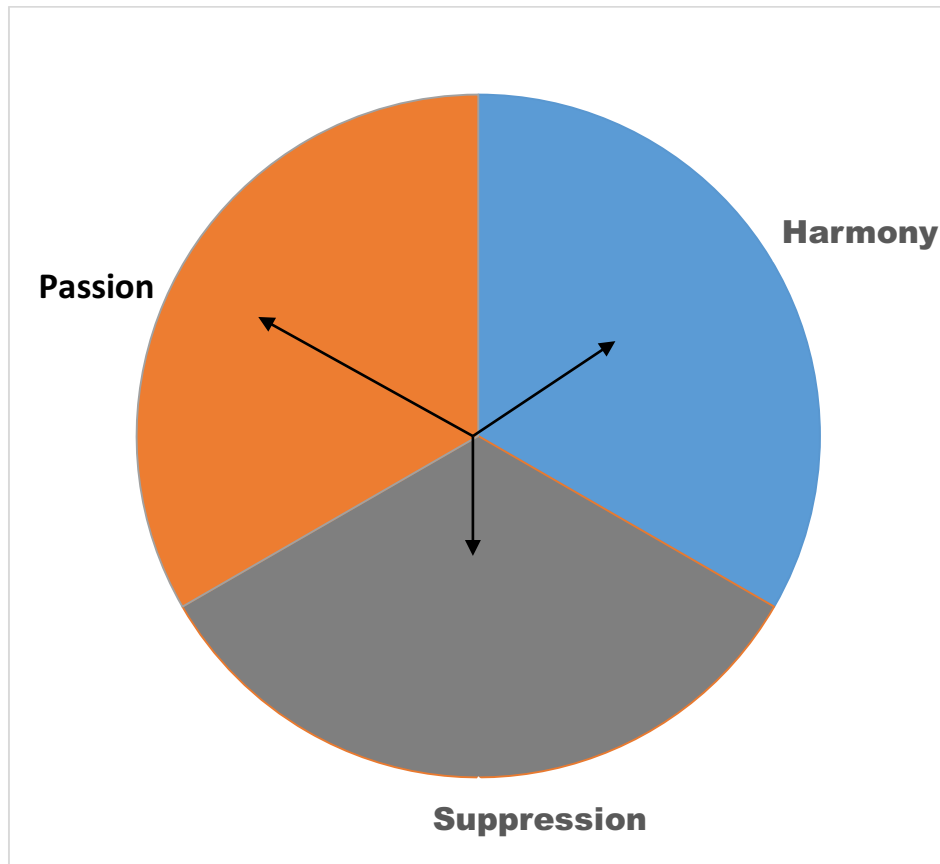

Figure 5: Sector arrows.

Sector arrows are important for policy analysis. It is also important to ensure robustness of sector arrows as much as possible. The robustness is necessary for the policies to stand up under scrutiny and for them to be effective over time. There are two independent robustness checks.

(a) One way to ensure robustness, is re-examine our choices of indicators add more statistics, and make the length correction. Assuming that good statistics are available, or that the poor quality of the data will be overcome by adding more, we can continue to *add* statistics, until we discover that the length and direction of the sector arrows *stabilise*. They will do so under the *corrected* length.

This is a meta-statistical step in the method, and is inspired by the notion of convergence in Bayesian statistics. Assuming the luxury of reliable and independent statistics, we should find relative stability in the sector arrows after the corrections in length to the indicator arrows have been made. The relative stability is what brings robustness to policy in a very real mathematical and statistical sense. Moreover, this is another reason for adding the arrows not all together, but adding the indicator arrows *within* each third separately. We thereby treat each general quality independently of the others. An advantage of thinking in terms of convergence and stability is that we can stop collecting statistics in a sector when we have stability! When we have stability in a sector arrow, we have a good statistical reason to think that we have collected enough statistics to show the degree of the general quality as it is manifested by the institution. But careful here about the word 'stability'.

'Stability' does not mean fixed. As already mentioned, institutions change, both for internal reasons and because the surrounding context changes. We can up-date the statistics individually, and watch the arrow change over time. What is important is that *before* we do that we have reached a robust arrow at a given time. However, we have to be careful for another reason: there might be something that we overlooked, that pulls the arrow significantly in direction or length. To fend from overlooking something important in our table of statistics, we have a second robustness check.

> (b) The second way to ensure robustness is to consult a wider audience to decide on the sector, angle and length of the indicator arrows and to solicit for new indicators. Thus, we consult not only people in the administration of an institution, but also others involved *in* the institution, or those affected *by* the institution. This might again force us to re-consider the naming of the statistic, and might require that we make two or more indicator arrows out of one.

If we meet the two robustness checks, this will ensure longevity of policy in the sense of having fewer surprises. There is always the possibility of error. All we can do is reasonably minimize it. But also consider the political aspect of consulting widely. If they have the time, and if they care about an institution or its effects, then people *like* to be consulted in the policy decision process. They feel then that their voice has been heard, and that their opinion counts. Having had their say, then by also witnessing the concerns of others they will understand and accept the final policy better. Consultation and inclusiveness also helps to ensure longevity of policy.

> (iii) Draw a triangle linking the three heads of the sector arrows as in figure 6.

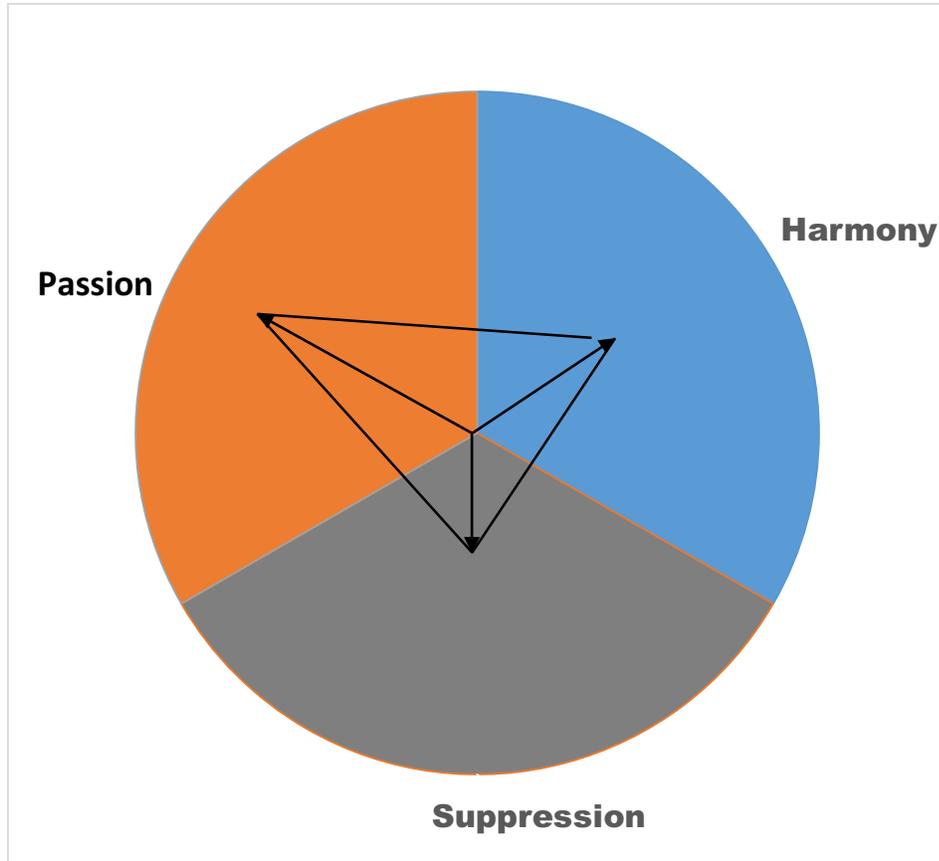

Figure 6: triangle uniting the heads of the sector arrows.

(iv)     Erase the sector arrows. See figure 7.

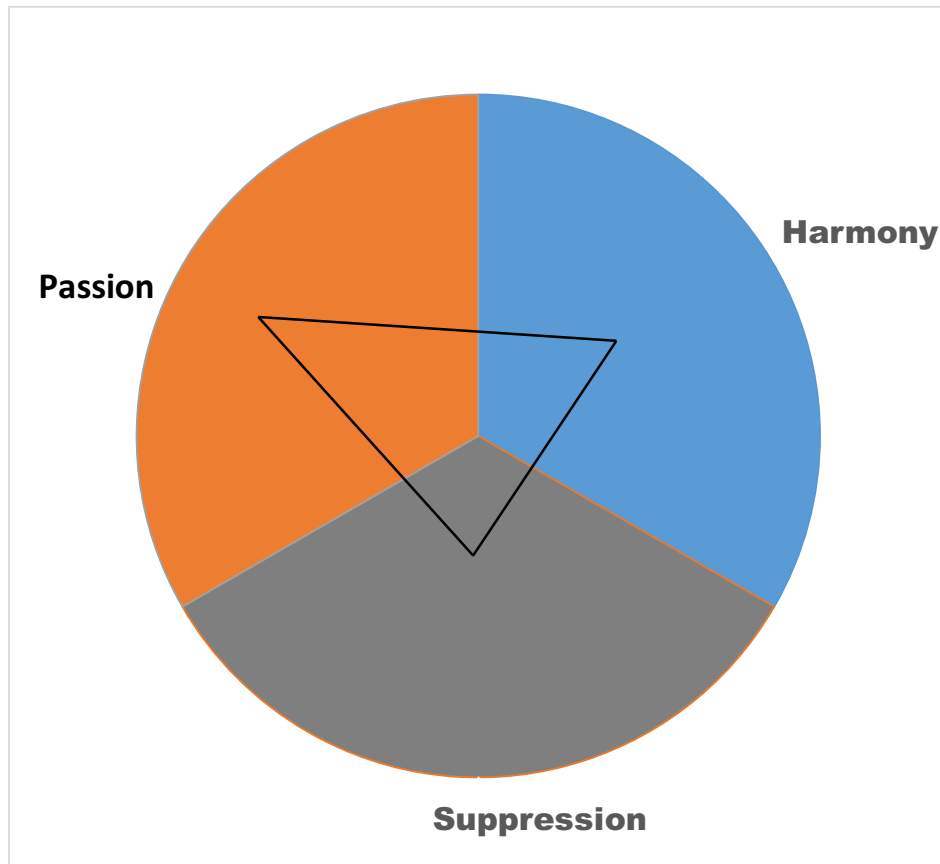

Figure 7: triangle without sector arrows.

(v)　　Find the centre of the triangle.[9] See figure 8.

---

[9] There are three different centres of triangles. The relevant one for us is the one that picks out the average points in the triangle. To find this draw a line from a corner to meet the opposite side with a right angle. The centre is where the three such lines intersect.

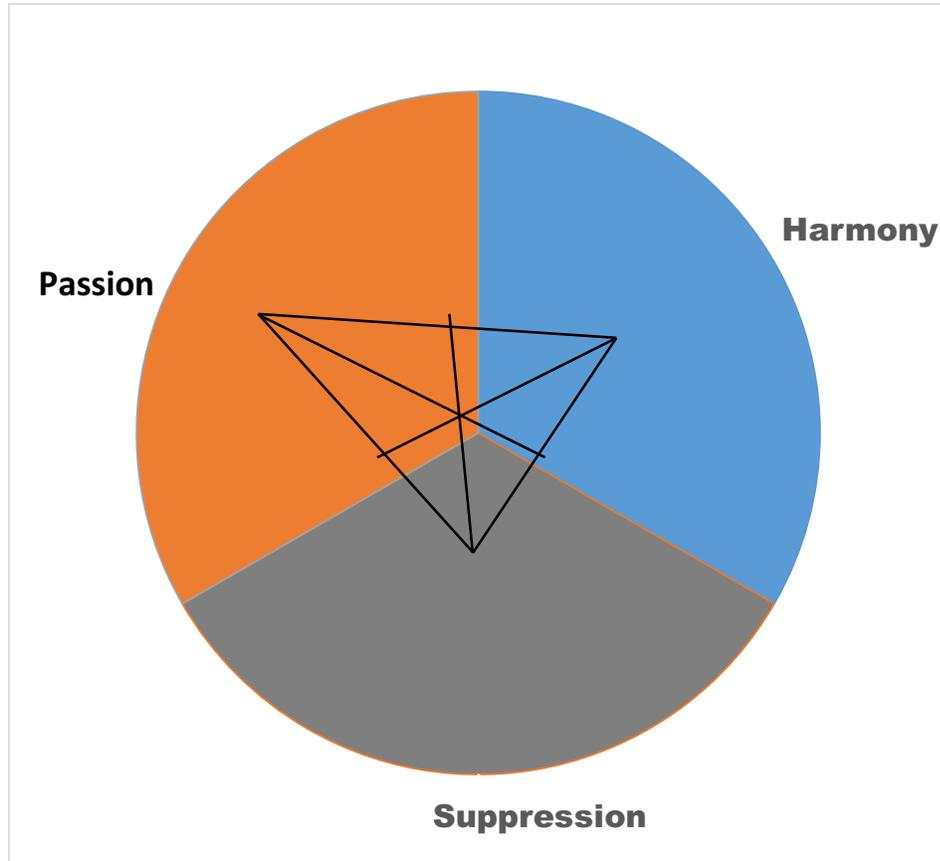

Figure 8: finding the centre of the triangle.

(vi) Draw an arrow from the centre of the circle to the centre of the triangle. See figure 19.

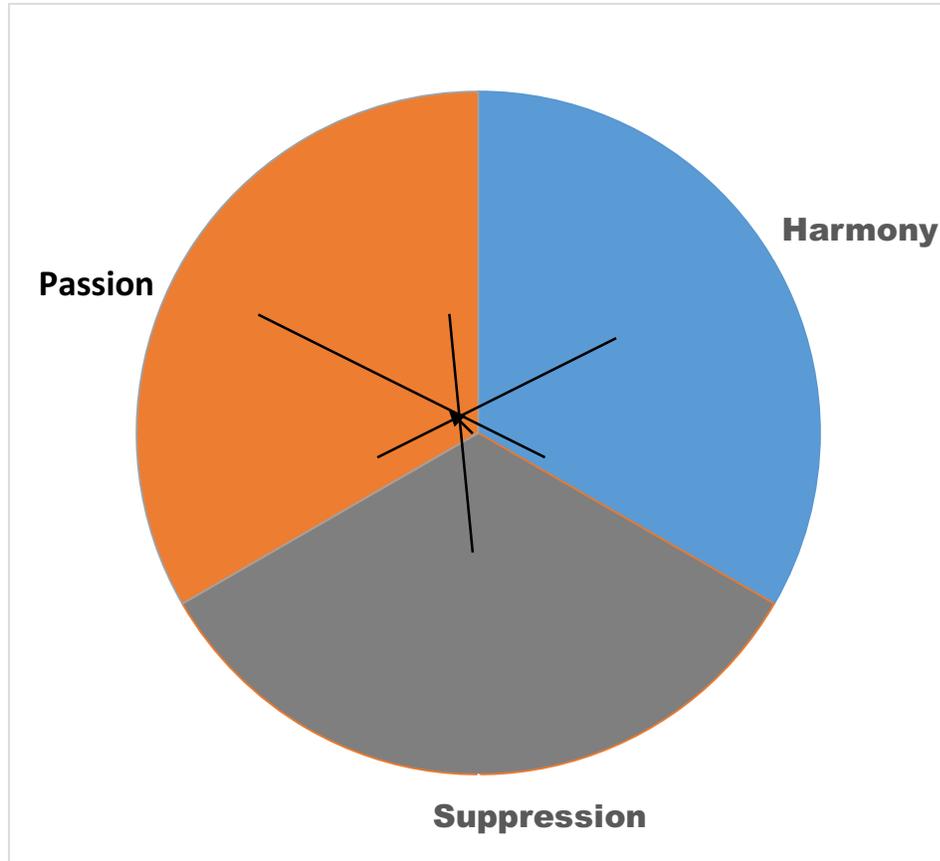

Figure 9: arrow from centre of the circle to the centre of the triangle.

(vii)   Erase the lines used to find the centre of the triangle. See figure 9.

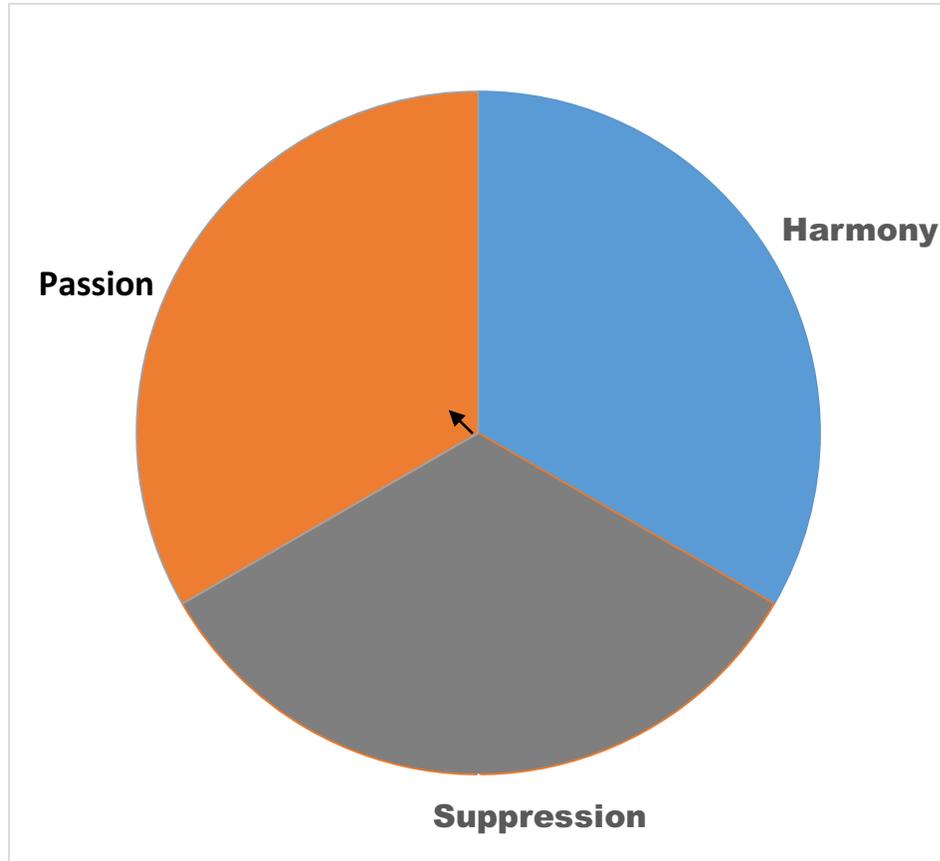

Figure 10: arrow from centre of the circle to the centre of the triangle.

**(viii)** We do our correction trick again in order to make the representation more perspicuous. If the arrow from the centre of the circle to the centre of the triangle is less than half the radius, then take its exponent. If it is half the length, do nothing. If it is more than half the length, then take its logarithm base 10. Draw the corrected arrow on the circle. This is our final arrow given the particular table of statistical data we started with. See figure 11.

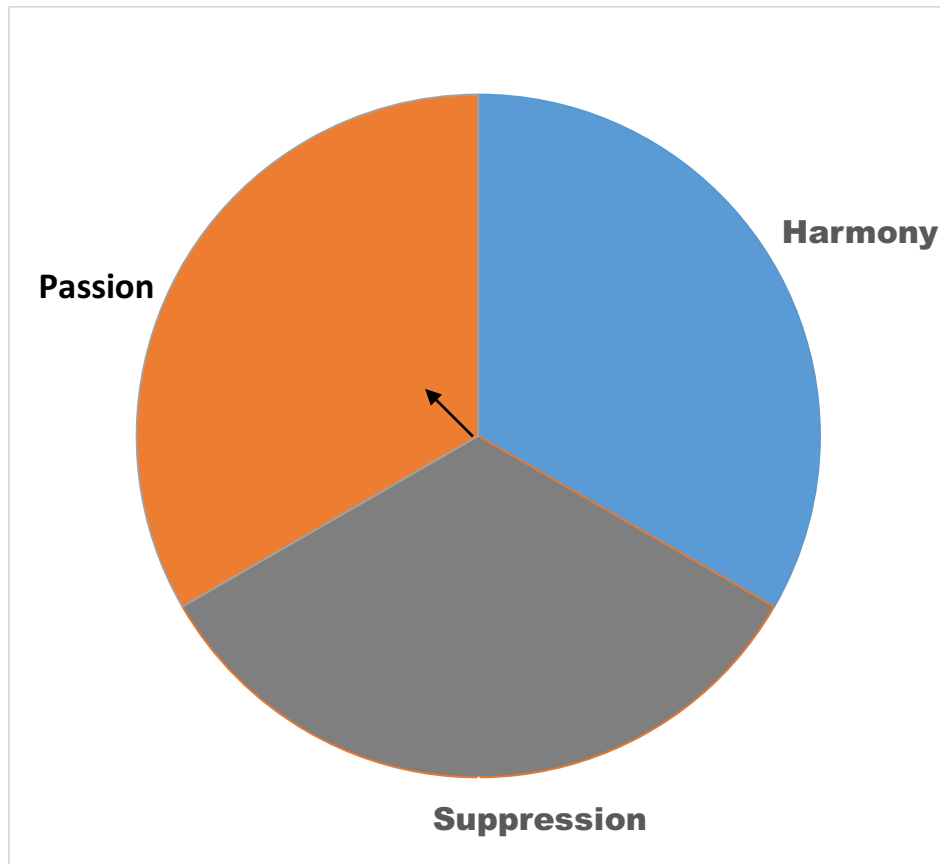

Figure 11: final arrow (with corrected length).

This is the *final* arrow *given* the particular indicators we chose, together with the direction and length we thought represented that indicator. It is a compass reading, giving us an orientation in terms of the three general qualities.

To read the final arrow we would see that the institution in question is in the passion sector. The arrow is fairly short, so the institution is fairly balanced between the three qualities. We might be quite satisfied with this result, depending on our culture, the mandate of the institution and its context. However, now we should watch out because the danger with passion is that it can turn into suppression. To fend from this, we would watch that the indicators in suppression do not become longer, that those in passion and harmony become longer. Which ones are easier to control or modify by policy will depend on the institution and its setting. But before we decide on policy we have a little more work to do.

We can now make new policies based on the final arrow, change existing policies, criticise policies and justify policies. We make the new policies, changes, criticisms and justifications based upon recovering the story we told in developing the indicator arrows. As was remarked in the previous section we can perform these exercises in a superficial manner or in a deeper manner. If our arrow is robust, then the margin between superficial and deep treatment diminishes.

## 3. Adapting the Compass to Align it with Ecological Economic Thinking

We now consider our institution from the perspective of ecological economics. This has three separate elements, the first conceptual, the second conceptual-normative and the third is ethically-normative.

The conceptual element is that we think that economic activity, the econo-sphere is a subset of / is dependent upon, society. The socio-sphere and society is a subset of / is dependent upon, the physical and biological environment: the eco-sphere. We also consider that we live in a world subject to at least the first two laws of thermodynamics; that the planet we live on is limited in space and resources; that some of our practices have harmful and irreversible consequences on the natural environment; that there are values other than value in exchange, such as intrinsic value and use value.

The conceptual-normative element is that as ecological economists, we want the world to be a certain way. We then criticise institutions on the basis of the extent to which they align with that vision. What is the vision? The we should make policies that *bring us closer towards* living within the natural flows: of heat from the sun, and heat dissipated from the earth, the flows of water, the flows of air. We wean ourselves from using up our fund of low entropy. The rate at which we wean ourselves and approach sustainability within the natural flows is culturally sensitive. We, decide for example on a *culturally acceptable rate* of entropy production, of pollution, of disruption of flows and of biodiversity loss. We actively engage in trying to slow these rates. We recognise three sorts of value: value in exchange, use value and intrinsic value.

The ethically-normative element is one step of generality up. We consciously endorse the above conceptual-normative element as a matter of moral principle.

To conform to the three elements, we make a more elaborate construction.

(ix)   Make tables and policy compasses for each sphere separately: one representing the relationship of the institution with the eco-sphere, one for the relationship of the institution to the socio-sphere and one representing the relationship of the institution to the econo-sphere. We now have nine sectors, three in each of three spheres. See figure 12.

| Table for *econo*-sphere | | | | | Table for *socio*-sphere | | | | | Table for *eco*-sphere | | | | |
|---|---|---|---|---|---|---|---|---|---|---|---|---|---|---|
| General quality | Name and qualifications | Angle | Length | Corrected length | General quality | Name and qualifications | Angle | Length | Corrected length | General quality | Name and qualifications | Angle | Length | Corrected length |
| Har | Steady wages | 20 | .5 | .15 | Har | general happiness | 30 | .6 | .2 | Har | Biodiversity of ecosystem | 30 | .2 | .06 |
| Har | Regular turnover of merchandise | 60 | .9 | .3 | Har | Regular work | 40 | .9 | .3 | Har | Stable mammal population | 90 | .6 | .2 |
| Har | books are balanced | 60 | .4 | .12 | Har | Regular use by members | 20 | .9 | .3 | Har | Predictability of rain | 100 | ..2 | .06 |
| Supp | debt | 160 | .6 | .2 | Supp | People suffering from depression | 130 | .2 | .06 | Supp | Mercury pollution in water | 200 | .9 | .3 |
| Supp | Cutting back | 180 | .9 | .3 | Supp | Days of sick-leave by employees | 200 | .8 | .27 | Supp | Unabsorbed carbon monoxide and dioxide | 190 | .6 | .2 |
| Supp | bankruptcy | 160 | .0 | .0 | Supp | Graffiti (incidences of) | 220 | .3 | .1 | Supp | Fossil fuels burned | 220 | .9 | .3 |
| Pass | sudden profits | 250 | .5 | .15 | Pass | Festive days | 300 | .6 | .2 | Pass | Acreage of crops | 300 | .6 | .2 |
| Pass | Earning spikes | 300 | .6 | .2 | Pass | Days of exhibition | 300 | .9 | .3 | Pass | Acreage of cattle | 250 | .9 | .3 |
| Pass | Re-structuring human resources | 350 | .6 | .2 | Pass | Reports in media | 320 | .9 | .3 | Pass | Acreage of private gardens | 330 | .2 | .06 |

Figure 12: Three tables, one for each sphere.

Start with the economy table. Economic value, is value in exchange. The time-frame is short, since exchange value changes quickly. The extent of the value is indicated by the length of arrows. Economic institutional *harmony* is indicated by steady wages, regular turn-over, the books balancing and so on. Economic institutional *passion* is indicated by large profits, surprise booms or spikes, changes in the economic structure. Economic *suppression* is indicated by debt, having to 'down-size', bankruptcy.

Social value of an object, or institution is value in use. The time-frame is longer than for the econo-sphere. Social institutional *harmony* is indicated by general contentment, regularity of work and use of the institution, lack of conflict with the society in which the institution is couched. Social institutional *passion* is indicated by festivals, exhibitions, provocative art work, sensational successes covered in the press. *Suppression* is marked by depression, disease, and violence.

Environmental value is intrinsic or existential value. The time-frame is long. Environmental institutional *harmony* is indicated by the state of nature without humans. There are flows and cycles and these blend and combine to show a steady-state of the environment and individual ecologies in the long term. Environmental *passion* is had when humans intervene. We control nature, we 'enhance' it and shape it. So indicators would include quantities and qualities of manicured gardens, indoor plants and farmland.

We have to always be careful with the passion quality because it can easily tip an institution into suppression. Environmental institutional *suppression* concerns: pollution of water, soil and air, covering soil by buildings or cement, waste, and especially entropic measures such as use of non-renewable resources.

For the purposes of ecological economics, we make the following general recommendations for the indicators in all the tables. Consider that we live in a world subject to at least the first two laws of thermodynamics, so some indicators have got to reflect the entropy production of the institution; that the planet we live on is limited in space and resources, so when we use up the natural space, there is a debt to pay. The general idea is to follow Georgescu-Roegen (1971) and think in terms of the fund-flow model of the natural environment. We have a fund of low entropy that we can use up quickly or more slowly. This includes non-renewable (or too slowly renewed) resources: coal, oil, gas. We want to use these up as slowly as possible – *aiming towards* the rate of replacement, although we know that it might be impossible on balance against social unrest, to *achieve* the use of slowly replaced resources *at or below* the rate of replacement. There will be more said on this in a moment. Following Georgescu-Roegen again, we note that there is a flow of heat from the sun and dissipation of heat from the Earth. Within this flow (over which we have slow control through the emission of greenhouse gasses) we respect the rate at which natural resources such as: food, fibre, wood, replenish themselves naturally. Another important sort of indicator comes from the thought that some of our practices have harmful and irreversible consequences on the natural environment. Here we use the metaphor of health of an organism and with some changes in the metaphor, apply it to whole eco-systems. Rapport (2012) has developed a number of these indicators. Most of them concern natural ecosystems, but they are stressed by human activity in the form of pollution. So the third, related, general idea is to think in terms of pollution, waste and clean-up. Once pollution has been released into the environment it starts to damage the natural eco-systems. There are limits over which the natural systems cannot cope and they become 'unhealthy' or 'collapse'. What replaces them is a *less healthy* natural eco-system. (Rapport and Maffi 2010). Some pollutants are dispersed quickly and others slowly, the effect of mixes between them is rarely known. The rates of dispersal should be respected and we should monitor the health of eco-systems to warn us of natural limits, in the case of accumulation or mixing of pollutants. Thus, the statistics falling under these three general ideas are what to use to find the relevant statistics for the eco-sphere.

To serve the ideas of ecological economics, of respecting the natural environment, we *would like* our institution arrows to be in the sector of Harmony. So as policy analyst or consultant using the ecological economics compass we advise that policy brings the arrow of the institution into the harmony sector. If it is already there, then carry on as before and address concerns in the socio-sphere or the economy-sphere!

Some examples of relevant statistics for the several spheres might help. Say the institution is a library. With respect to the institution's relationship to the environment, under 'suppressive' we might think of surface area of the building, pollution, energy used by the building. Under passionate we might wonder if there are any projects that help the environment – special exhibitions or displays, a bee-keeping project on the roof of the library, a special greenhouse space for reading. If all of these are absent, then the arrows would have no length. We might then want to think about how to have the environment passionate arrows have some length! In the Harmony third, we would consider the green space outside, or plants indoors, or recycling facilities and so on.

For creating our compass for the socio-sphere, under the quality of suppressive we might think of controls for taking out books, oppressive security measures, or lack of fresh air in the building (disease and fatigue inducing). In the passionate third we could consider the statistics concerning exciting exhibitions, a series of talks or media coverage. In Harmony, we would consider the steady adherence and participation in the library's facilities by the public, the facility with which they obtain material, the smooth running of the service in general. As for economic indicators the suppressive costs are rent, cleaning, repairs. Harmony might be looked at in terms of regular membership fees, steady acquisition of books or electronic material, payment of regular staff. Under passionate, the library might make money from a special exhibition, or hire an artist to come in and paint the walls in some exciting way, we might acquire new equipment that will attract new members.

The conceptual-normative element of ecological economics is represented by the sub-set relationship of the three circles. We think that the environment has to be given priority over society, and the latter has priority over economic considerations. This will now be represented by the relative *sizes of the circles*, representing the three tables. See figure 13.[10]

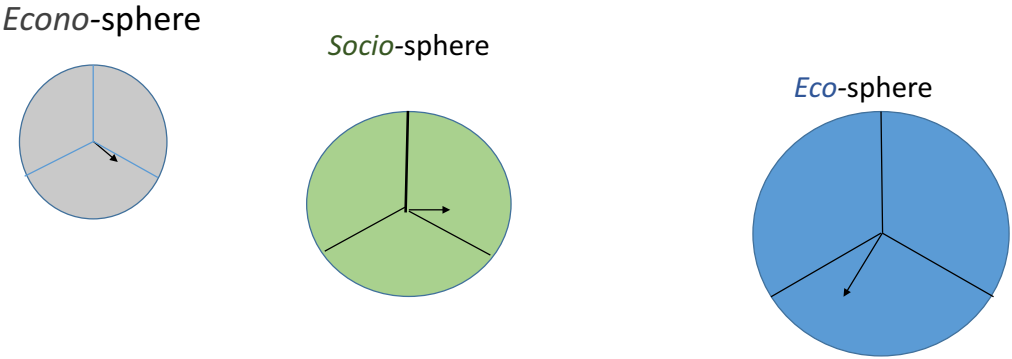

Figure 13: Three Spheres.

(x) We place the spheres concentrically in the following proportions: the radius of the eco-sphere is the longest. The radius of the econo-sphere is half the length of the eco-sphere, and the length of the radius of the socio-sphere is exactly half way between the two. See figure 14.[11]

---

[10] We do not label the three qualities, but they are still there.
[11] We erase the lines indicating the three quality sectors, because that is not what is important in this part of the construction. They are still there but in the background.

## Super-impose the three spheres

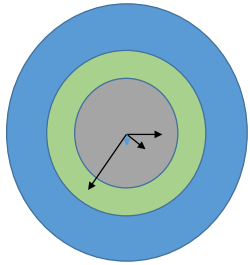

Figure 14: Ecological Economics Compass

The relative length of the radii of the three circles reflects how much more important we think, for example, that the eco-sphere is with respect to the socio-sphere. To fend from confusion, remember that in the table the length of the indicator arrows is given as a proportion of the length of the radius of *that* circle. The corrected length is then the logarithm base two of the length divided by the number of indicators in that sector. Thus, the calculations ensure that the eco-sphere arrow has more influence than the other two. Look after the natural environment first, then look after society, and lastly look after the economy. We aggregate the three final arrows for the spheres by simple vector addition, as in step (iii).

(xi)   We draw our three 'final' arrows, head to tail as in step (iii) with the first one's tail starting at the centre of the circle. Erase the circumference of the socio-sphere and the econo-sphere and the labels for the spheres. Return to the general qualities. See figure 15.

# Return to the qualitiative compass, erase the spheres, add the vectors

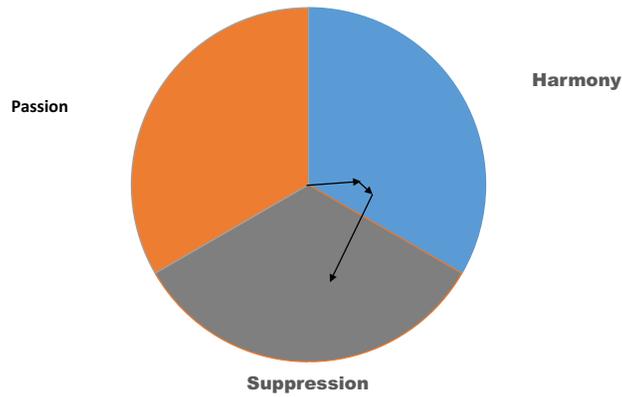

Figure 15: Ecological Economics Compass, re-introducing the qualities.

(xiv) Add the three 'final' arrows tail to head, and draw an arrow from the centre of the circle to the head of the last arrow. See figure 16.

# Draw an arrow from the centre of the circle to the end of the added vectors

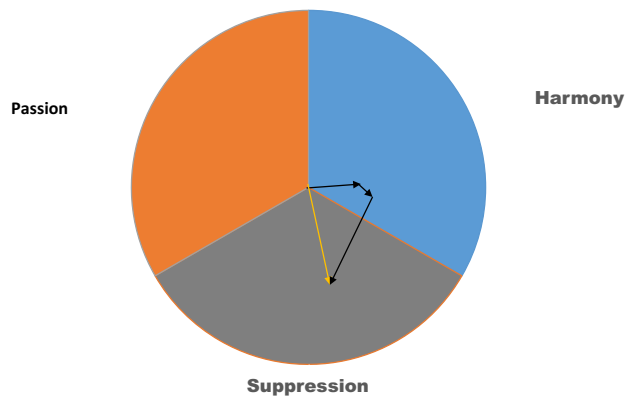

Figure 16: Vector addition of arrows from the three spheres.

(xii) Erase the indicator arrows. This is the final arrow for the institution, from the perspective of ecological economics. See figure 17.

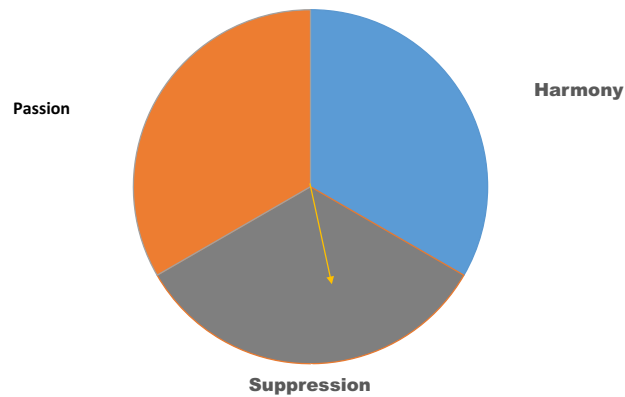

Figure 17: Final arrow.

This compass is constructed using the thinking behind ecological economics. If the environment is doing well, then we need not take care of it, and we can concentrate on society or the economy or both. Because the eco-sphere arrows are more influential over the position and length of the final arrow, it is the statistics concerning the environment that we will heed most in our policy decisions – unless the eco-sphere is doing well.

Returning to the three *gunas* or the three general qualities, for the purposes of getting along with the environment, we would like our final institutional arrows to be in the quality of *sattva*. Even an institution that is normally suppressive, such as a prison might have its final arrow in *sattva* – if healthy gardens are created and maintained, waste is well-managed and so on!

## 4. Sustainability and the Policy Compass

It is politically in *vogue* to claim that an institution is "sustainable". Such a claim is almost empty when we consider the number of different definitions we might have for the word 'sustainability'. An industry might be deemed 'unsustainable' just because it is not financially solvent; or, we might be concerned with sustaining a certain standard of living; or we might be concerned with the stability of an ecological system. The word 'sustainability' was used to refer to the natural environment, but has been appropriated by business and government because of the new, fashionable and strongly positive connotation. Since it is a vague but positive term, its meaning is seldom made explicit by business or government.

Ecological economists use the word in the older sense. This is to reflect the scientifically established reality of how it is that the environment is doing, and to acknowledge our dependence on that environment. We have to ensure that the eco-sphere is sustained and not collapsing. It is only if this is sustained, that we can think of sustaining society (within the bounds and context of the natural environment). This is important for government, other public institutions, many non-profit organisations and many NGOs (non-governmental organisations); but, at least for the ecological economist, government ought to take very seriously the natural environment in which, and from which, the society lives. Lastly, economic activity should not be 'sustained' at the 'cost' of society (stimulating social break-down). Thus, the ethical element is what plays into the notion of sustainability for the ecological economist. Similarly, a business institution which claims sustainability in the ecological economist's sense would take seriously both the social aspects internal to the institution, the external aspects: the society in which it is couched and the natural environment. Thus, I put forward the claim that

> Claim 2: for the ecological economist, an institution is sustainable iff the ecologial-economics policy compass's final arrow is in the third of harmony.

'Sustainability' is here taken to set an ethical and normative standard; reflecting what it is we want to sustain and what we are willing to sacrifice. Under ethical-normativity, we want for the natural environment to remain relatively stable. In very basic terms we need to ensure that water, the air and the soil are natural and harmonious. It is only then that the bio-sphere can continue in a healthy manner.

We all know that increasing entropy damages the environment, and therefore, entropy production is one of the obvious choices for representing the suppressive with respect to the environment. Another environmentally suppression indicator is pollution. Insofar as we are willing to sacrifice the environment to our social or economic ambitions, we think less and less in alignment with ecological economists. Let us be quite clear. There is the scientific aspect of ecological economics and the more ethical aspect. It is this distinction that has been drawn out in the exercise of constructing our ecological-economic policy compass.

## 6. Philosophical Remarks and Conclusion

As noted in the introduction, we can construct several aggregated arrows, representing change in an institution over time, or for comparing institutions to each other. The construction is sensitive work due to our making qualitative decisions, as per section two. But it is not impossible.

> Remark 1: The first philosophical remark is that our analysis can be shallow or deep.

We can address an 'outlier' statistic in our table (step one of the methodology), where an 'outlier' statistic is one that influences the length and direction of the aggregated arrow disproportionately. We can simply erase it and replace it with a 'better behaved' statistic, or we can change direction and length (by changing context or further qualifiers and by changing scale of the outlier arrow). We can also change the relative proportions of the three spheres. Of course with such gerrymandering, relative 'objectivity' is then sacrificed, since rather than deciding on the direction and length *independently* of the other indicator arrows, we do so *with respect to* the other indicator arrows and in particular with respect to the final arrow and its representation on the ecological economic three spheres. This is trickery, but might be enough to justify a policy in the short term or to a gullible audience. The robustness checks counter the gerrymandering, so it is with a less robust compass that gerrymandering is easier. Remember that there are almost no surprise outlier arrows left in a robust compass.

A deeper analysis can be made by working out how to change the reality behind the length and direction of the various arrows. This can be done through changes in policy that address that statistic. For example, if an alarming number of people are dying of lung cancer, and this is statistically linked to smoking heavily, then we design policies to encourage general lung health, and decrease smoking.

An even deeper analysis involves looking at the statistics more thoroughly: re-examining the context and culture that give weight to the decisions concerning degree and length. We can also look for the underlying causes of the statistics within the predominant or secondary third (we look back to the sector arrows), or by looking at the whole, hence the holistic aspect of the analysis. Returning to the smoking example, we might notice that smoking is considered to be 'cool' in some sub-cultures, so we can try to counter that image in the broader media. We might also notice that it is associated with rebellion or disquiet. In this case, we might want to look for means of making people feel more at ease and in tune with society.

> Remark 2: The second philosophical remark is that the compass design is not *ad hoc* with respect to the three qualities.

We might be tempted to change the compass by changing the qualities. We can do this in two ways. One is by adding or subtracting general qualities. The other is by changing the three general qualities to another three general qualities.

Let us start with the number three. We could add more general qualities. This is counter-recommended for the following reasons. The three qualities are conceptually orthogonal to each other. That is, they were chosen to be in some sense incommensurable. Yet every institution displays one of these predominantly. They are not strict opposites. That is why there are three and not four. We cannot make a policy compass by dividing the circle into four since this would suggest that the two opposing quarters are opposites. Thus, to mathematically represent the orthogonality/ conceptual incommensurability of four general qualities, one would need three dimensions. Our circle compass on paper would have to be a sphere. This is by no means mathematically difficult, or difficult to represent with computer interface, since we can show it being rotated around, but such representations are of limited use because they are difficult *for us* to see and understand, despite the computer interface. Remember that, here we are interested in making policy decisions based on a *simple* final representation, and the simplicity is what would be sacrificed. Similarly, for five general qualities, we would need four dimensions and so on. For reasons of ease of understanding and reading the representation on screen or paper, three is an optimal number for general qualities.

We could *change* the three general qualities to three others. For example, in deciding on the general quality and suitability of a piece of fabric, we might want to compare durability, beauty, insulation. Durability would be indicated by how easy it is to tear or wear out the fabric; beauty by possible patterns imprinted, shininess, pliability; and insulation by togs, noise penetration, wind penetration and so on. A piece of fabric is not an institution, so we have widened the scope in terms of what the compass arrow exercise can be applied to.

Changing the content of the qualities is possible, but remember where they come from. The three qualities come from the Indian philosophical traditions: Hinduism, Jain and Buddhism. The original words in Sanskrit are: *sattva* for harmony, *raja* for passion, and *tamas* for suppression. A guru will assess and advise someone based on which of the three qualities predominates in that person and which needs more emphasis. While *raja* is exciting and glamourous, it can easily tip into *tamas*, as we try to hold on to the excitement and glory. The path of wisdom is to follow a more *sattvic* life, tending a little towards *raja*. According to Kumar (2007) from whom I draw inspiration, *sattva* is also the better direction for living in harmony with the environment, which is one of the reasons I chose to translate *sattva* as harmony.

We could do both: change the qualities and *reduce* the number of general qualities and just have two polar opposites, say, 'good' and 'bad'. This is simple, but loses subtlety and sacrifices too much of the complexity of the situations in which we try to make difficult policy decisions. 'Good' and 'bad' are relative terms and are sensitive to context and time. Forgetting these sensitivities is what leads to *poor* policy decisions.

There might well be times when we think that it is worthwhile to compromise the eco-sphere, for example to prevent immediate war, or under political pressure from other countries. We would then reduce the relative size of the eco-sphere. Kozo Mayumi (2001, 45) puts it very nicely: as a society, or institution, we decide on a culturally acceptable rate of entropy production. Following Friend (2017, 101), if we add "per unit of consumption", then we have a culturally acceptable measure of ecological *efficiency*. Both are important for the purposes of ecological economics. We might also allow some institutions more leeway in compromising the environment than others. These are decisions it is better to make consciously than unconsciously – again adding depth to the analysis, creating a solid defence of a policy and informing changes in policy.

Remark 3: There are checks for robustness of the sector arrows.

By 'robustness' we mean that the sector arrows are quite stable – adding more indicators does not alter the angle or length of the sector arrow very significantly – there are no *undiscovered* 'outlier arrows' – ones that significantly change the direction or length of the sector arrow, and therefore the final arrow.

To ensure robustness, or accuracy of the information, we have two checks. (a) We look for convergence on the length and position of the sector arrows, (b) we consult more people involved with the institution when we make the table. We ask them if there is an indicator that we overlooked, and consult them for assessment of length and position. It might turn out that there is a strong disagreement, and then we make two separate indicators. The idea behind wide consultation is that the sector arrows will be more robust if we consult people lower down in the institution and people outside the institution and who are affected by it. Through participation in the exercise of developing the indicator arrows, then will understand and better accept the final policy decisions.

It is not always possible to ensure robustness. If we lack robustness, then we make policy decisions that address the lack of data, the poor quality of the data or lack of consultation. In this way we understand the *limitations* of the compass exercise.

Remark 4: Even when we do not have a robust compass, and enough data, we still learn valuable lessons from the exercise of constructing the compass.[12]

It is not always straightforward or possible to construct a compass that passes the robustness checks. In these cases, we develop policy recommendations to get better statistics! We can choose them, and name them very carefully in advance. It is with the material we have, and under constraints, that we do the best we can. We then have *an* explanation as to why we made a particular policy decision, and a *partial* justification for the policies, but the explanation *includes* discussing the quantity and quality of the statistics, so it is a more abstract and technical explanation and justification.

More important, we make our policy recommendations with a degree of hesitancy. By trying to make a policy compass we have a good sense of why we are hesitating, and we can share this information with those concerned. They can then be vigilant about the success of the policy and understand the importance of having a good suite of reliable statistics. Moreover, we have a pretty good idea of how to change the policy if we later learn new statistics that influence the position or length of the non-robust arrow. So we can anticipate. After all, when we make policy decisions, we know in advance that we do so with imperfect information, that the context changes, that values change, but we make the decisions despite this. With the use of a policy compass, even a non-robust one, we do *better* than just take a guess, trusting our instinct, or making the decision on the basis of a monetary calculation to maximise profit.

---

[12] Nicole McLernon drew my attention to this, and together we explored the merits of using a non-robust compass.